\documentstyle[pre,aps,epsf]{revtex}
\begin{document}
\title{Longevity of orders is related to the longevity of their constituent genera \\
rather than genus richness}
\author{Stefan Bornholdt$^1$, Kim Sneppen$^2$, and Hildegard Westphal$^3$}
\address{$^{(1)}$Institut f\"ur Theoretische Physik, Universit\"at Bremen, 28359 Bremen, Germany  \\
$^{(2)}$Niels Bohr Institute, Blegdamsvej 17, 2100 Copenhagen, Denmark \\
$^{(3)}$Fachbereich Geowissenschaften, Universit\"at Bremen, 28359 Bremen, Germany}
\maketitle
\begin{abstract}
Longevity of a taxonomic group is an important issue in understanding the dynamics of evolution.
In this respect a key observation is that genera, families, or orders can each be assigned a characteristic average lifetime (Van Valen, L. 1973, Evolutionary Theory 1, 1-30). Using the fossil marine animal genera database (Sepkoski, J.J.Jr. 2002, A Compendium of Fossil Marine Animal Genera, Bull. Am. Paleontol. 363, 563 pp.) we here examine key determinants for robustness of a higher taxonomic group in terms of the characteristics of its constituents. We find insignificant correlation between the size of an order and its stability against extinction, whereas we observe amazingly large correlation between the lifetime of an order and the lifetime of its constituent genera.
\end{abstract}
\bigskip 
  
\section{Introduction}
Extinction and existence times of taxonomic groups open for a window into the interplay between robustness and diversity by presenting a large number of examples of how history at lower levels of the taxonomic hierarchy influence higher levels. One interesting insight into this interplay is the Red Queen hypothesis of Van Valen \cite{1} stating numerous case studies that existence times of a family within a specified order was exponentially distributed; a fact that indicated the randomness of extinctions as well as the distinctive different lifetimes between the different orders. This widely recognized difference in characteristic extinction rates between higher-level taxa goes back to Simpson's ``Tempo and Mode in Evolution''  \cite{2}. Concerning the actual patterns and causes of extinction events in particular, Raup has made interesting observations, including ``kill curves'' and fractal patterns of extinctions, as well as possible periodicity in the large-scale pattern of extinction  \cite{3,4}. Such periodicity has later been studied, e.g., by Kirchner  \cite{5}.

The pronounced taxonomic overturn in the fossil record of the Phanerozoic suggests that intrinsic differences exist between the taxonomic groups in their evolutionary properties  \cite{6}. Contrary, studies have been suggesting that evolution and extinction to some extent can be simulated as a random branching processes and accordingly described without regards to differences between taxonomic groups  \cite{7,8}; This hypothesis has been criticized by others  \cite{9,10,11,12,13}, and we will also here present evidence against such a simplification.

Also we will discuss correlation between stability of a taxon, and of its constituents, a correlation which would be found in models based on branching processes  \cite{14}. Empirically, this question has been explored by Sepkoski  \cite{15}, Sepkoski and Miller  \cite{16}, and Gilinsky  \cite{17} among others. In this paper we will use the Sepkoski database of fossil marine animal genera  \cite{18}, where origination times and extinction times of about 37,000 genera are recorded. Recent studies of this database focused on rates of originations and extinctions and on the time pattern of activity in the fossil record (see, e.g., refs.  \cite{5},  \cite{19}). We here study other aspect of the data, namely the statistics of origination times and extinction times and their relationship with taxonomic levels.

\section{Results}
Figure 1 shows the database as a ``duration matrix'', i.e. a compressed matrix format, where each point denotes a time interval with a specific origination time and a specific subsequent extinction time, representing the lifetime interval of one or more genera who share the common fate of origination and extinction in the same respective stratigraphic intervals (duration matrix). Thus, the vertical distance from a point to the diagonal is the duration time, or lifetime, of the corresponding genus or genera (whereby clearly an order cannot have a shorter lifetime than the longest-lived genus in that order). As seen from the high density of dots close to the diagonal, most genera have short lifetimes compared to the Phanerozoic. The figure also quantifies how large extinction events separate the matrix into a near block-diagonal form as, e.g., the Permian/Triassic extinction, that 250 million years before present (Ma) separated Paleozoic life from Mesozoic life. Although most genera have short lifetimes, the dots far above the diagonal in Figure 1 show that some genera have long lifetimes and a few even very long lifetimes. Finally, the modern elements of the Paleozoic, Mesozoic, and Cenozoic marine fauna line up at the top edge of the plot. The plot not only illustrates the intensity, e.g., of the ÒBig FiveÓ mass extinctions, it also shows the composition of the fauna elements going extinct with respect to origination time.

\begin{figure}
\epsfxsize=18cm
\begin{center}
\epsffile{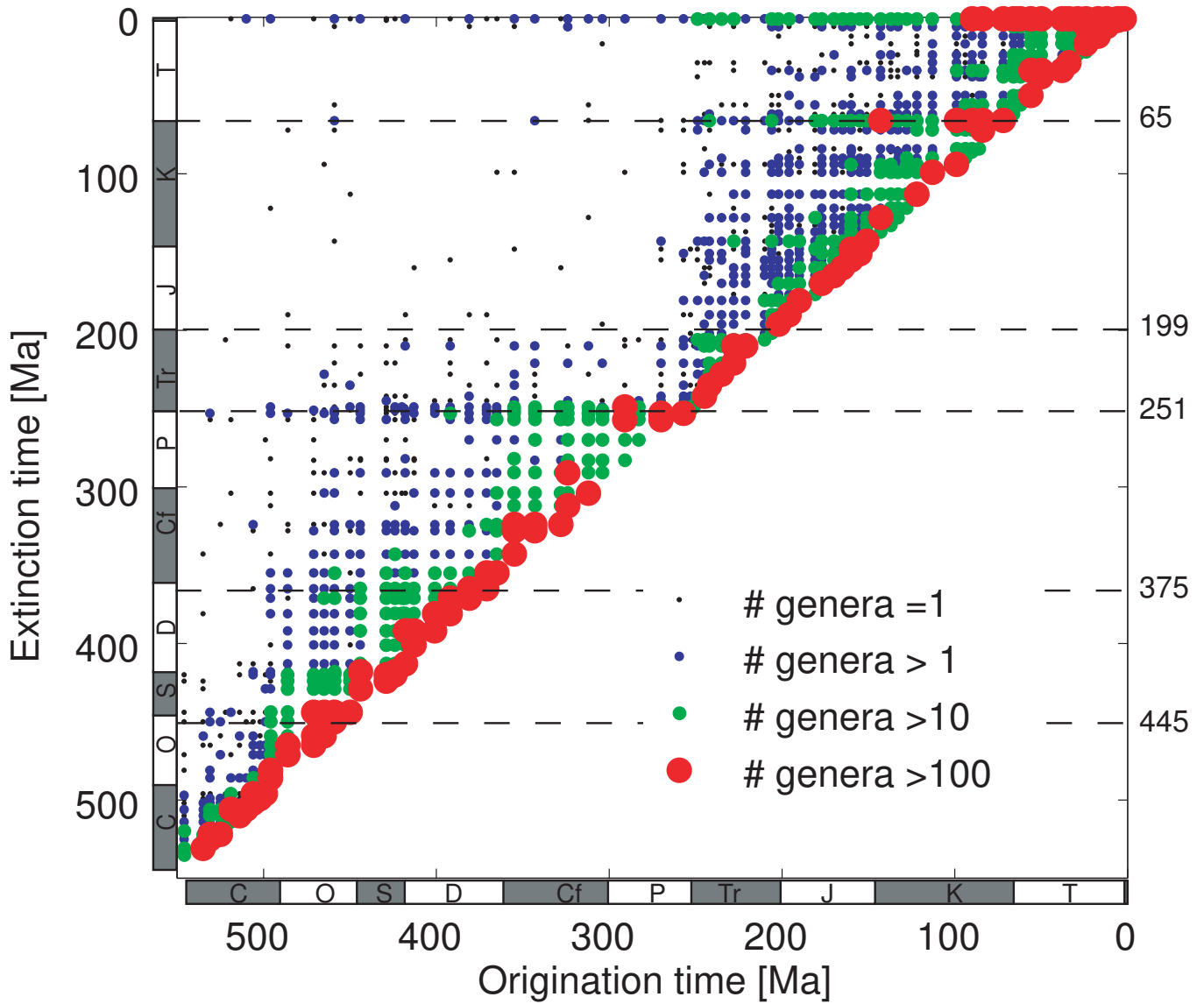} 
\end{center}
\caption{ Origination and extinction of approximately 37,000 marine animal genera in the Phanerozoic. Points marking extinctions at $t = 0$ indicate genera living up to the Recent. The vertical distance from a point to the diagonal measures the corresponding genus lifetime. Notice the collapse of many points close to the diagonal, reflecting the fact that most genera exist less than the overall genera average of about 27 Myr (million years). }
\end{figure}
Figure 2 shows the lifetime data ordered by taxonomic groups, following the ordering given in the Sepkoski database that is taken as approximating taxonomic relationship. In Figure 2a each genus duration period is represented by a vertical line starting at origination and ending at the end of the geological stage where it was last observed. This presentation reveals correlations between fates of genera listed close to each other. The more closely listed genera often tend to share both origination and extinction time. Also the data appear clustered in the sense that whole groups often are going extinct together. The result is that lifetimes between related genera are strongly correlated, a fact that is illustrated in Figure 2b that gives the lifetimes of all the genera in the same ordering as Figure 2a. This plot highlights a huge variability in lifetimes, with most genera existing for a shorter time interval than the overall genera mean lifetime of 27 Myr. The figure strongly pinpoints that closely related genera have closely related life spans, and implies that survivorship is heritable.
\begin{figure}
\epsfxsize=14cm
\begin{center}
\epsffile{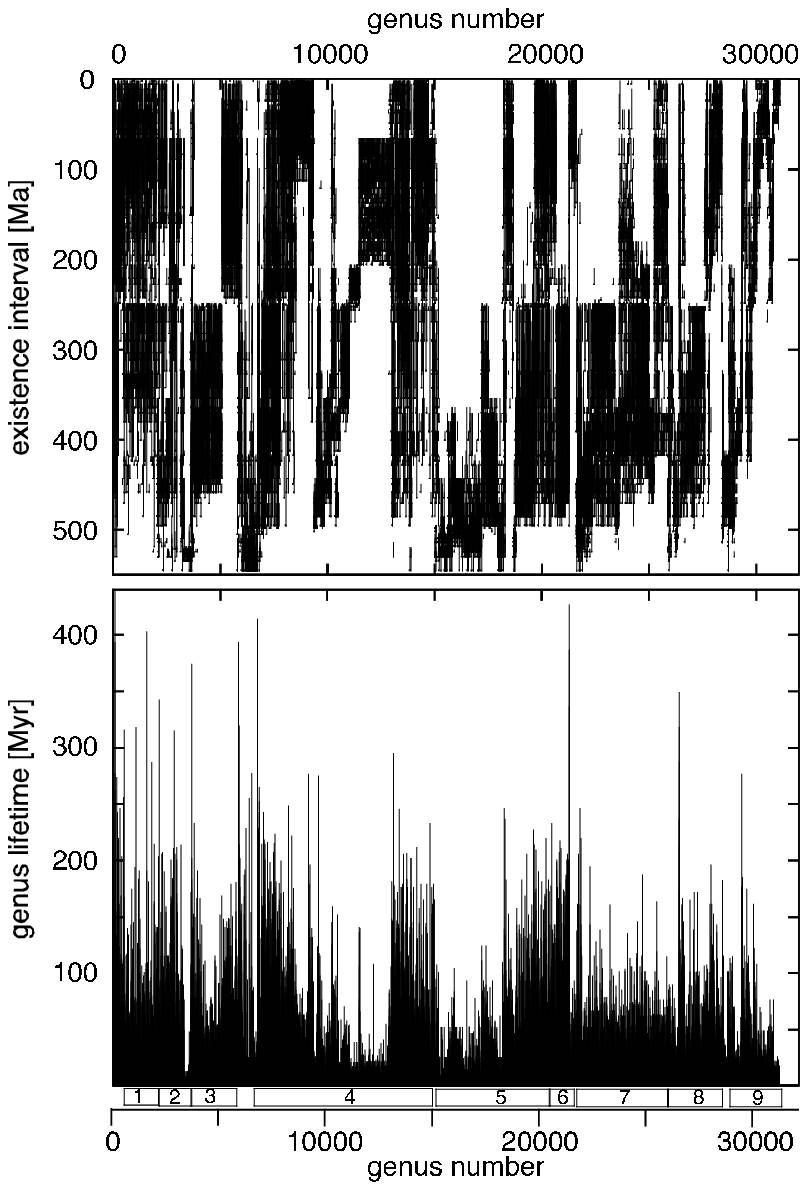}
\end{center}
\caption{Lifetime of genera represented by vertical lines, sorted by taxonomic groups as given by the ordering of the Sepkoski database. (a) Lifetime of each genus with respect to geological stages of origination and extinction. (b) Lifetime durations of genera in the same ordering as in (a) show that lifetimes between related genera are strongly correlated. Large phyla are marked in plot by numbers at the x-axis of (b): 1=Rhizopodea; 2=Porifera; 3=Cnidaria; 4=Mollusca; 5=Arthropoda; 6=Bryozoa; 7=Brachiopoda; 8=Echinodermata; 9=Chordata.}
\end{figure}

\begin{figure}
\epsfxsize=15cm
\begin{center}
\epsffile{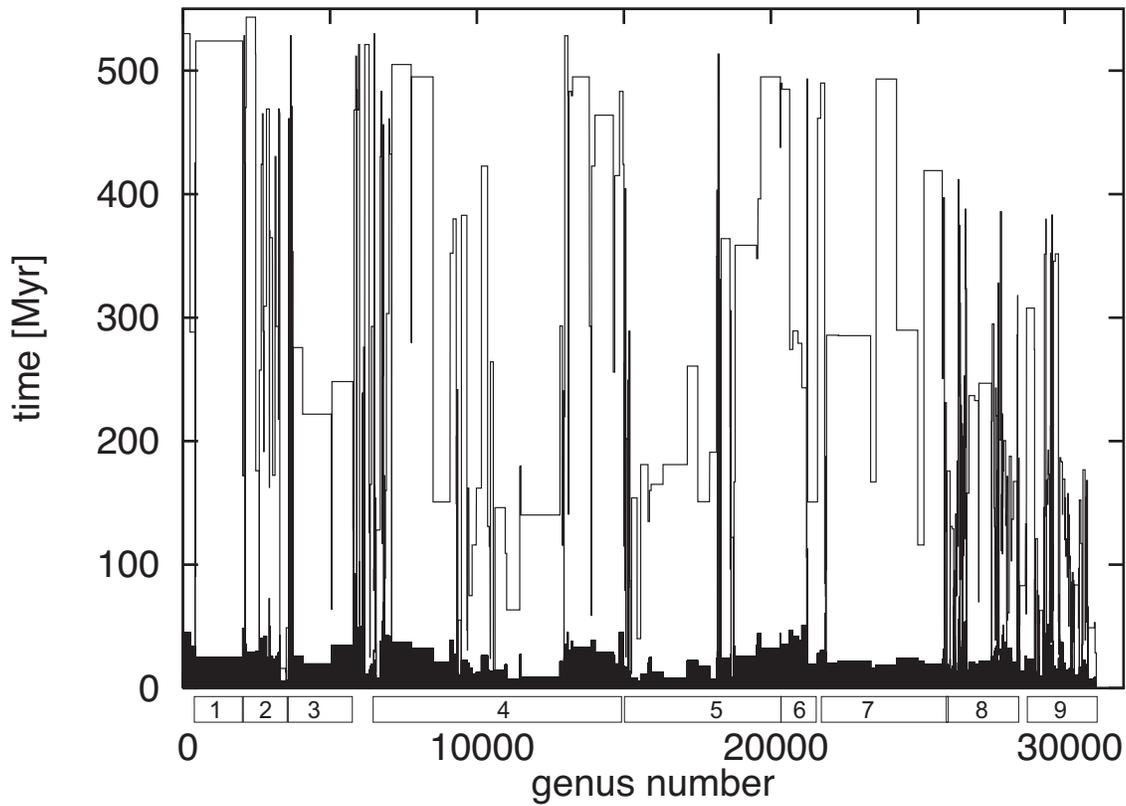}
\end{center}
\caption{Lifetime of orders (thin lines) and average genus lifetimes within orders (black shaded histogram). Orders with numerous genera are represented by wide blocks in the histogram. Ordering of genera as in Figure 2; phyla are marked as in Figure 2.}
\end{figure}
Figure 3 re-examines Figure 2 in terms of ordinal lifetimes and average genera lifetimes within orders, resulting in subdividing the data into blocks consisting of orders, containing between one and more than 1000 genera. 

the form $P(s)=1/s^{1.5}$. 
\begin{figure}
\epsfxsize=12cm
\begin{center}
\epsffile{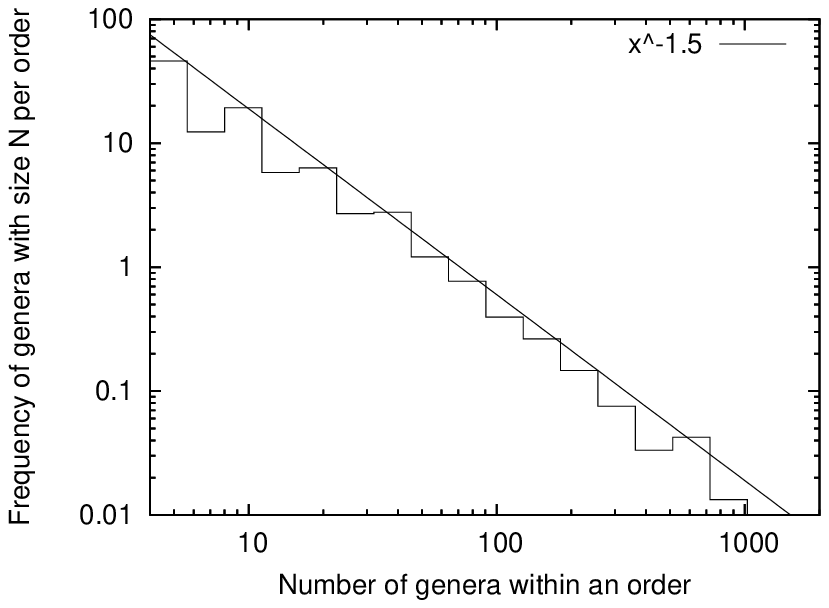}
\end{center}
\caption{
Distribution of order size as quantified by the integrated number of genera during the lifetime of the respective order. One observes a near-perfect power law; $1/size^{1.5}$.
}
\end{figure}
In Figure 4 we show the distribution of numbers of genera in the various orders, a plot that indeed displays a very broad distribution that can be amazingly well represented by a scale free distribution of the form $P(s)=1/s^{1.5}$. Such power laws for taxonomic groups have been previously reported by Yule  \cite{20} for genera size distribution in terms of number of extant species. In this regards we obtain the $1/s^{1.5}$ scaling both for the accumulated number of genera within the total existence of the orders (shown in the Fig), as well for as for the number of genera in orders at any given time cohort (with upper cut of at about 100). Returning to Figure 3, the plot shows a box-like distribution, each box representing an order with a width equal to the number of constituent genera, and with a height given by the average lifetimes of the genera in the order. Figure 3 thus demonstrates that even large orders might have very different average genus lifetimes, with differences which are much larger than random expectation would predict. In Figure 3 we also show the total lifetimes of the orders. The two lines runs roughly parallel indicating that there is correlations between the two properties. In contrast, there is no systematic tendency that big orders, represented by broad boxes, are longer lived than smaller orders (narrow boxes).

To elucidate the overall statistical properties of genera lifetimes, seen over the whole taxonomic record, we in Figure 5 investigate genera lifetime distributions in the subset of genera that went extinct (there are no extant genera in this analysis). 
\begin{figure}
\epsfxsize=18cm
\begin{center}
\epsffile{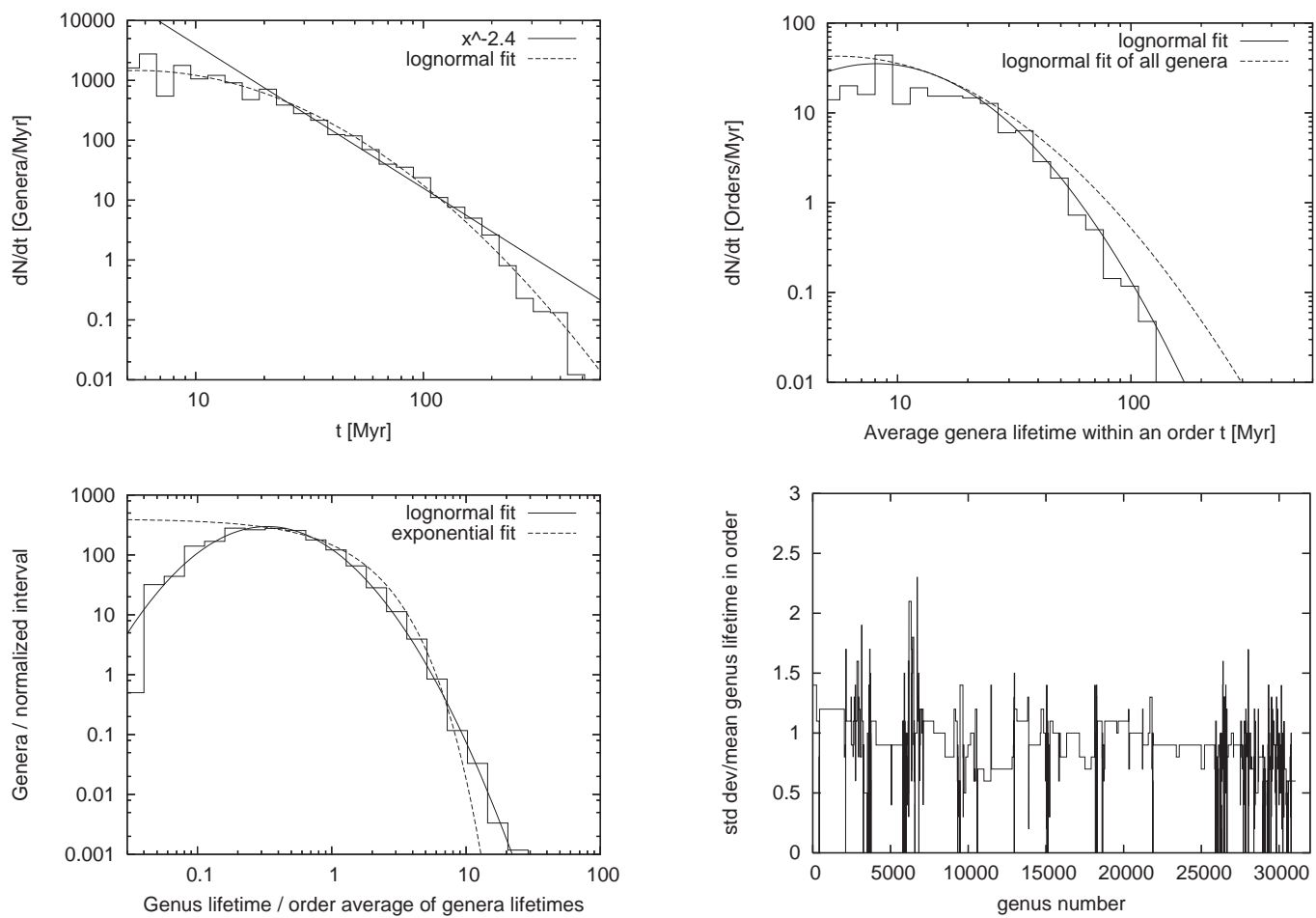}
\end{center}
\caption{
(a) Distribution of genera lifetimes throughout the Phanerozoic. (b) Distribution of average genus lifetimes within orders. (c) Collapsed plot of genera lifetimes inside orders, renormalized by their respective average lifetime within the order t order. (d) The ratio of spread to mean of genera existence times within respective orders. For comparison, this ratio equals 1.34 when considering the full dataset.
}
\end{figure}
Figure 5a demonstrates that genera lifetimes are indeed so broad, that they can be fitted by a $1/t^{2.4}$ power law from 20 to 200 Ma  \cite{21,22,23}. At larger time intervals the lifetime is limited by the length of fossil record, whereas genera with shorter lifetimes tend to be suppressed as they are more easily missed from observations. The genera lifetimes are also consistent with a log-normal distribution, as illustrated by the slightly bent dashed line in the figure, fitting well the overall distribution with mean lifetime of 27 Myr and with a standard variation of 36.1 Myr. Figure 5b shows the distribution of average genera times within orders from Figure 3. In accordance with the observation of the different heights of the boxes in Figure 3 one observes a quite broad distribution of these average genera lifetimes, ranging from a characteristic scale of a few Myr to 100 Myr: In fact these averages can also be fitted by a log normal distribution. This distribution is narrower than the overall distribution of genera lifetimes in Figure 5a, the difference reflecting the scatter of individual events within each order. Figure 5c finally rescales all genera lifetimes to the expected mean for their respective orders. The resulting distribution is so narrow that it can well be fitted by an exponential, that is only slightly wider than the exponential distribution Van Valen found in his study of a sample of taxonomic groups  \cite{1}. This is in spite of the fact that the numbers of genera within different orders vary over a large range, (from one to 1000 genera per order, in fact, as also seen later) Such an exponential distribution is consistent with a model where each genera have a characteristic stability sets by its heritable properties, and goes extinct solely due to a external random event that comes from outside, such as climate changes or co-evolutionary changes in its environment. Fig 5d finally decomposes Fig. 5c into the normalized spread to mean of lifetimes for each order, $\sigma(t)/<t>$. For an exponential distribution, this number equals 1, whereas a larger $\sigma(t)/<t>$ pinpoints a distribution of lifetimes which is broader than exponential. In particular, considering lifetimes of all genera we have $\sigma(t)/<t>=1.34$. Figure 5d indeed shows that most orders have well defined lifetimes, given by a single exponential. The only significant exception is Foraminiferida from genus number 1000 to genus number 3000.

We now return to our prime observation of Figure 3, namely that lifetimes of orders do not correlate with their size but rather correlate with the average lifetime of genera within them. To substantiate this observation we confine our study to orders that (1) do not exist any more, (2) contain more than four genera, and (3) have no gaps in the record. This reduced sample of 159 orders with in total 18,534 genera is subsequently examined in Figure 6.
\begin{figure}
\epsfxsize=18cm
\begin{center}
\epsffile{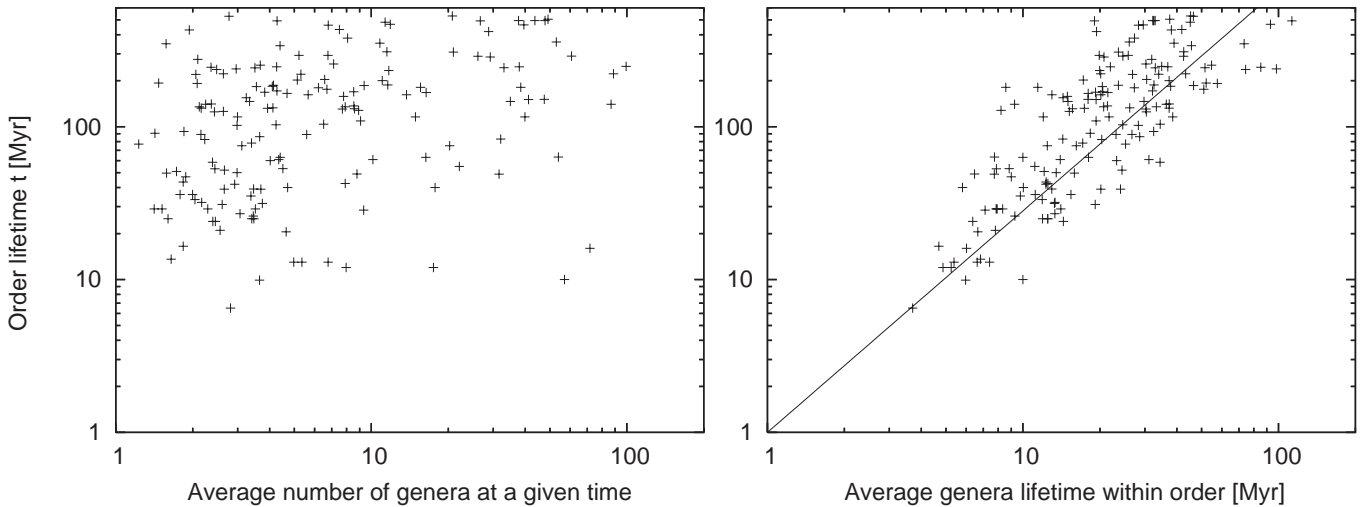}
\end{center}
\caption{
(a) Existence time of orders versus average number of genera at a given time during the respective order history. We only examine orders which have more than four genera, have no gaps in their record, and which are extinct now. (b) Existence times for same sample of orders as in a) are here shown as function of the average existence times for genera composing the respective order.  The straight line $f(t) = t^{-1.45}$ represent a least square fit to the data, and is shown to emphasize that the order lifetime grows more than proportional with the characteristic time of the genera composing the order.
}
\end{figure}
 Figure 6a shows a scatter plot of the lifetimes of orders and the average number of constituent genera of these orders during their duration; i.e. it shows how well one could predict the future fate of an order based on how numerous the genera in that order are at a given time. The figure, however, reveals that there is weak or no correlation between order lifetime and the average number of its constituent genera. That is, if we calculate the average order lifetime for genera with less than three genera it is 111 Myr, whereas it is 235 Myr (only about a factor of two more) for orders with more than 30 genera per time slice (well over one order of magnitude more genera). Nevertheless, order lifetime is correlated with the average lifetime of its constituent genera (Fig. 6b). Orders lifetime tends to scale with $t^{1.5}$, where $t$ is the average genera lifetime. This scaling for example means that with genera, which are four times longer-lived, the order tends to be eight times longer-lived.

\section{Discussion}

Species richness as well as geographic range, local abundance, reproductive mode, body size and inferred generation time, trophic strategy, and life habit that have been put forward as significant factors for survivorship during Òbackground extinctionÓ times  \cite{24,25}. Our findings contrast with the notion that numerous and geographically widespread taxa are more robust against extinction  \cite{26}. Such an intuitively appealing notion is indeed the mathematical consequence of any model where extinctions of individual genera are random and independent from each other.

The fact that no detectable correlations exist thus in itself pinpoints that similar genera share similar fates as confirmed by the strong correlation between neighboring genera (Fig. 2a). Another unexpected observation is the fact that the lifetime of an order increases more than linearly with the lifetime of its constituent genera. That is, one would trivially expect that the lifetime of an order increases proportionally to life of its constituents, i.e. if every constituent exists double as long the whole lifetime of the group would be doubled. Even a less dramatic model as suggested by Pigolotti et al.\  \cite{8} where genera replicate or go extinct with equal probability and thereby generate an overall power law distribution of life times of higher taxa, also predicts at least a proportional growth of order lifetime with the number of its constituent genera per time cohort. Thus the lack of correlations between order robustness and size (Fig 6a) pinpoints some coordination of extinctions at least between closely related genera. In this regards we have also found no correlation between size of order at last stage before extinction, and how log time the order existed: Even very large orders at the peak of their existence occasionally undergoes sudden extinction. In fact 10\% of the orders had a larger genera count than their average at their last stage.

Another interesting facet of our analysis is Figure 6b which shows that the lifetime of the whole grows disproportionally fast with the lifetime of its constituents. In fact, given that order lifetimes of less than a stage length and more than the geological record cannot be observed, means that the order lifetime roughly scales with the genera lifetime $t$ as $t^\gamma$, where $\gamma$ is between $1.5$ and $2$. That is, the line in Figure 6b was obtained from a least square fit giving $t^{1.5}$, whereas a null model where we regenerate order lifetimes as function of $t$ given by $t^\gamma$ and with the upper and lower cut of 5 and 540 Myr gives best fit for $\gamma=2$. Obviously the individual order lifetimes scatters around this mean expectancy, reflecting the stochasticity of events that causes any particular order to disappear.

A secondary finding confirming Van ValenÕs  \cite{1} original Red Queen hypothesis is the exponential distribution of genera lifetimes within a given order (Fig. 5c). As also noticed by Van Valen, an exponential distribution of lifetimes implies that the extinction of a genus is a random event, happening with a rate given by some properties that are similar for all genera within an order. In contrast, a broader than exponential distribution obtained for the full data set of lifetimes shows that genera from different orders have different intrinsic stabilities to whatever could cause their extinction (Fig. 5a) From this we conclude that all genera within an order have closely related (although not necessarily identical) stabilities, whereas genera in total have a much broader distribution of intrinsic stabilities. Note that whereas the intrinsic stability against extinction is already defined at origination, the actual extinction event exhibits the statistics of a random event that only depends on external factors, including competition from other species. Observable intrinsic stabilities may result, e.g., from heritability of geographic range  \cite{27} that in turn affects survival during the major mass extinctions in the geologic past  \cite{28}; and a general negative relationship between geographical range and extinction rate on species level as has been demonstrated for different taxonomic groups  \cite{24,27,28,29,30,31}.

Our good fits of genera lifetimes at different taxonomic levels with log-normal distributions suggest a heritable stability, with could be represented by products of random numbers, the more similar the genera the similar their respective numbers in the product. Such modeling is well known from physics of complex systems, including fragmentation phenomena and turbulence heuristically modeled already by Kolmogorov and by Filippov  \cite{32,33,34}.

Evolution of life has involved strongly varying rates of origination and extinction with a general tendency of decreasing turnover over the Phanerozoic \cite{35,36,37}. The effect of the overall decline in turnover rates over the Phanerozoic as described e.g. by Flessa and Jablonski  \cite{26} is not reflected in our analyses of ordinal longevity. We divided the 159 orders that fulfilled the criteria for being in Figure 6 into subsets of 77 orders that existed only during the Paleozoic (before the Permian-Triassic extinction), and 43 orders that existed only during the Mesozoic to Cenozoic. In both cases we again observe no significant increase in order lifetimes when comparing orders with less than three genera in average, to orders with more than 30 genera in average during their existence, matching the observation for the full phanerozoic shown in Fig. 6a. Similarly, we reanalyzed the correlation with average genera lifetimes, and at least the Paleozoic orders confirmed our faster than linear scaling from Figure 6b. The statistics of the Mesozoic to Cenozoic orders were too weak to be conclusive. Thus, overall the behavior of the Paleozoic taxa compared to the Mesozoic and Cenozoic taxa does not show any significantly different behavior. This means that there is no dependency in behavior on the geological time when the order existed.
\bigskip

To summarize from our study we conclude that:
\begin{enumerate}
\item There is substantial variation among higher taxa (orders) in characteristic genus durations.
\item Stability of higher taxa is predicted by stability of lower taxa. Put another way, orders that are longer lived tend to consist of genera that are longer lived.
\item Longevity of orders is not well predicted by the number of constituent genera.
\end{enumerate}

Further study will be needed to explore to what extent lifeÕs tendency to specialize, i.e., to create numerous variants over a single theme, is optimizing local fitness gains maybe even on the cost of longterm robustness against extinction.

\section{Data and Methods}
We use the last version of Sepkoski's database of fossil marine animal genera as published in ref.  \cite{18}. The effect of the incompleteness and bias of the fossil record  \cite{38,39,40} is minimized by studying this large database.

Genera originations (approximated by first appearances) and extinctions (approximated by last appearances) are translated to absolute ages using 98 distinct time intervals (on stage level where practicable; and otherwise merging stages) derived from the current IUGS Geochronology  \cite{41} and the Harland stratigraphic time scale  \cite{42}. To check sensitivity of our analysis to the exact choice of stratigraphic timescale, we cross-checked with a second implementation of time intervals based entirely on the Harland stratigraphic time scale \cite{42} and found similar results. For our analyses, we define the origination time of a genus as the beginning of the stratigraphic interval where this genus first occurred, and the extinction time point as the end of the stratigraphic interval where it last occurred. The results of our study do not depend sensitively on this convention.

\acknowledgements
The authors thank Sergei Maslov for stimulating discussions and comments. Thanks are extended to Eric Holman and David Raup for thorough reviews of an earlier version of this paper. Support from the Danish National Research Foundation through the Center Models of Life at the Niels Bohr Institute as well as the Deutsche Forschungsgemeinschaft is gratefully acknowledged.

\vfill\eject
\end{document}